\documentclass[letterpaper,11pt]{article}

\usepackage{url}
\usepackage{apalike}
\usepackage{fullpage}
\usepackage{graphicx}
\usepackage{hyperref}
\usepackage{subfigure}

\newcommand{\xf}[1]{Figure~\ref{#1}}

\newcommand{\xs}[1]{Section~\ref{#1}}

\newcommand{\lucidL}[1]{{$\mathit{Lucid}$}($L$) }

		{}

\def\myvert{\raise 2.27pt \hbox{\vrule depth 0pt height 8pt width 0.2mm}}
\def\myarrow{\hspace*{0.43mm}%
             \raise 2.29pt\hbox{\vrule depth 0pt height 8pt width 0.16mm}%
             \hspace*{-0.32mm}%
             $\longrightarrow$
             \ %
             }

\date{April 25, 2009}

\begin{document}

\title{The Role of Computer Graphics in Documentary Film Production\footnote{The abstract of this work has been accepted for presentation at FSAC 2009, Ottawa, Ontario, Canada \cite{role-cg-docu-film-presentation-2009}}}

\author{Miao Song\\
Graduate School\\
Concordia University, Montreal, Canada\\
\url{m_song@cse.concordia.ca}\\
}

\maketitle

\section*{Abstract}

We discuss a topic on the role of computer graphics in the production
of documentaries, which is often ignored in favor of other topics.
Typically, except for some rare occasions, documentary producers
and computer scientists or digital artists that do computer graphics
are relatively far apart in their domains and rarely intercommunicate
to have a joint production; yet it happens, and perhaps more so
in the present and the future.

We attempt to classify the documentaries on the amount and techniques of computer
graphics used for documentaries. We come up with the initial
categories such as ``plain'' (no graphics), ``in-between'', ``all-out'' -- nearly 100\%
of the documentary consisting of computer-generated imagery. Computer
graphics can be used to enhance the scenery, fill in the gaps in the
missing story-line pieces, or animate between scenes. It can incorporate
stereoscopic effects for higher viewer impression as well as interactivity
aspects. It can also be used simply in old archived image and film restoration.

\section*{Fair use of imagery}

The author declares the use of screen captures and the related imagery
in this academic work are for the illustrative reasons under the fair
use rationale (similarly to Wikipedia) where such work may still be
under copyright and the author does not make any claims of ownership
or authorship of the said images. The screen captures and the like
are attributed in the text and are necessary to illustrate the points
presented in this academic research paper.
A certain number of images from the cited documentary films are already
in the public domain or have compatible licensing.

\section{Introduction}
\label{sect:introduction}

There are several different types of computer graphics (CG) techniques used in
the modern documentary production, such as
computer-generated 2D and 3D images (CGI), computer animation, and even ``manipulation''
of the data generated by the computer, which is also called computer-human
interaction~\cite{wiki:computer-graphics,wiki:human-computer-interaction}.
Computer graphics has been widely used in physics, biology, video game industry,
education, and military matters. It has also been used in advertising,
music video, motion picture, and television such as international network promos
for CBC, e.g. shown in \xf{fig:cbc} produced in 1985,
and in the feature film {\em Jurassic Park} (1993), e.g. shown in
\xf{fig:jurassic_kitchen_raptor}~\cite{OSU-CG-history,wiki:jurassic-parks}.
In the past, computer graphics seemed only to be associated with big budget production;
therefore, the production, such as most of documentary film done in low budget
were rarely applied these new techniques in their production.
However, in recent years, more and more documentary films started to introduce
computer generated 2D and 3D images in a partially or fully animated
film.

No matter a hand-drawn animated documentary or computer-generated animated documentary,
they have certain aspects in common -- they are ``broadcast'' to the audience.
However, the interactivity between
human and computer generated images is a particular special characteristic related to computer
technology. Thus, there is ``interactive documentary'' -- a brand new genre,
which has various forms, such as 3D and web-based, where the audience can
participate in one way or another to as what is happening in the documentary.

\begin{figure*}[ht]
\hrule\vskip4pt
\begin{center}
	\subfigure[Canadian Broadcasting Promo]
	{\label{fig:cbc}
	\includegraphics[height=2.0in]{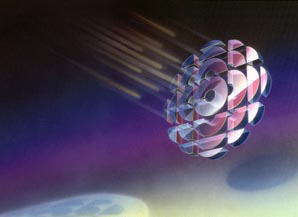}}
	\hspace{.3in}
	\subfigure[Jurassic Park's Raptor]
	{\label{fig:jurassic_kitchen_raptor}
	\includegraphics[height=2.0in]{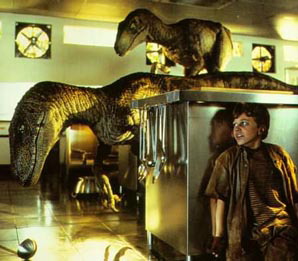}}
\caption{Some Example Graphics for Production}
\label{fig:cg-production-examples}
\end{center}
\hrule\vskip4pt
\end{figure*}

\subsection{Animation, Realism, and the Truth-Value}

``Truth-value'' and ``film realism'' are two often opposing topics debated in academic discussion
since 1990s~\cite{life-reproduced-drawings-2005}.
Some filmmakers try to enhance the realism of a film because some of the scenes in it are not ``real''.
Some other filmmakers say that there is no perfect truth in documentary film
because as soon as the film has been edited, some of the truth
has been hidden and is a subject to the opinion of the maker.
For some hybrid documentary films, some scenes have to be reconstructed.
Do the reconstructed scenes still have the truth value?
Compared to traditional documentary film, the scene
in an animated documentary does not ``exist'' at all. It is neither a
live-action camera footage nor a photograph, but is created by artists based on the
real world objects or based completely on their imagination.
How much truth-value will remain in an animated documentary film after reconstruction?
Does it exist there at all?
We always talk about how to make a film more realistic and increase its realism
in order to make the ``faked'' scene look more believable. Obviously, for such a film,
the truth-value is not comparable with the one that contains real footage and photographs.
Computer animation has a lot of similarities when compared to the traditional
animation. However, it is more advanced than the traditional one and more and more
adopted by the current film industry because of its efficiency, accuracy, lower cost, and more
advanced rendering effects, which can enhance the realism of a film more than traditional one does.
Even though computer graphics has been playing a significant role in formulating
new aesthetic grounds for both fiction and non fiction film, will this kind of film
abandon the most important concept, ``truth-value'', which is precious in documentary film?
How this new type of documentary film will affect the older audience and new generation audience?
How much could they believe CGI documentaries?
Through the discussion about this and other topics, we will reach to a
conclusion at the end in that regard.

\begin{quote}{\it ``Realism is not what animation is best at, instead, freer invention,
fantasy, and exaggerating reality are its privilege. The realer-than-real environment
follows the concept of {\em hyper realism}, which offers a completely
artificial environment as a representation of the real.''}~\cite{understanding-animation-1998}
\end{quote}

\noindent
Rowley in~\cite{life-reproduced-drawings-2005} talks about the quest
of traditional animated film (hand-drawn style) for a particular kind
of realism in order to succeed in feature film making, such as
``visual realism'', ``aural realism'', ``realism of motion'',
``narrative and character realism'', and ``social realism''.
We find these concepts are quite applicable to our discussion topic,
``computer animated graphics and the realism of documentary film''.
We borrow these terms for completely different perspective of analysis
in ``computer animated documentary'' instead of
``tradition animation in feature film'' discussion in Rowley's article.
We only talk about some of these concepts which have
a direct relationship with our work and mention others for completeness
in \xs{sect:cgi-and-realism}.

As shown, the biggest challenge for most of the documentary filmmakers is to capture truth.
Moreover, it is a difficult task for them to make a film when there is a shortage of footage.
There are various solutions to fill in the gaps, such as scene reconstruction,
photographs, texts according to historical events, and the needs from artists and the makers.
Some artists decide to use the photos to represent the past, such as
American filmmaker Ken Burns who uses the ``photographing of live-action material''~\cite{wiki:ken-burns}.
Some use either traditional animation or computer graphics, or both to reconstruct the scenes because
they feel it is more vivid and can make their scenes closer to the truth compared to photos.
Some explain the necessity to use animation in documentary to show the footage
that one can not capture in the real life, such as imagination, dream, hallucination, and memory.
A typical example such as in documentary {\em American Teen}
(2008)~\cite{paramount:american-teen,wiki:american-teen}, animation was
used for ``teen dream'' and ``teen imagination''.
In {\em Waltz with Bashir} (2008), an ``all-out'' animated documentary,
the animation was used to portray the memory of Ari Folman (the director of this film),
in the first Lebanon War twenty years after the war.
What types of computer graphics rather than animation
have been used in documentary production? We will discuss this further
in \xs{sect:visual-effects-types} and \xs{sect:example-mock-documentary}
with some examples of animated documentary film.

\subsection{Traditional Animation and Computer Animation in Documentaries}

\begin{figure*}[ht]
\hrule\vskip4pt
\begin{center}
	\subfigure[Example 1]
	{\label{fig:sinking-lusitania1}
	\includegraphics[height=2.0in]{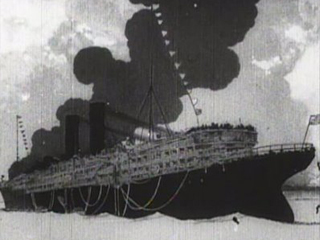}}
	\hspace{.3in}
	\subfigure[Example 2]
	{\label{fig:sinking-lusitania2}
	\includegraphics[height=2.0in]{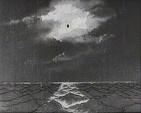}}
\caption{{\em The Sinking of the Lusitania} (1918) Example Screenshots}
\label{fig:sinking-lusitania}
\end{center}
\hrule\vskip4pt
\end{figure*}

The use of animation in documentary production is not new.
The traditional animated documentary can be traced back all the way
to 1918 by Winsor McCay in his 12-minute-long film
{\em The Sinking of the Lusitania} (1918)~\cite{sink-lusitanian-1918}.
It used animation to portray the 1915 ``sinking of RMS Lusitania after it was struck
by two torpedoes fired from a German U-boat''~\cite{wiki:sinking-lusitania}.
Quite obviously, there was no live-action footage was recorded when the event occurred.
The animation used in this silent documentary film for things such as the underwater
fish swimming and the explosion and sinking of Lusitania, some of which shown
in \xf{fig:sinking-lusitania}. The explaining texts were throughout
the film between the animated scenes. This is very traditional story
telling in silent film. The only difference is the animated scene replaced
the live-action footage.
Animation has been used in other educational and social guidance documentary
as films, such as {\em The Einstein Theory of Relativity} (1923). The purpose
to use animation in earlier times is because of the need to be able to
illustrate abstract concepts in mainly live-action examples of these
genres~\cite{truth-toons-documentary-1997,walking-life-animation-truth}.
In traditional animation, inbetweening, cell animation~\cite{TJ84}, and rotoscoping~\cite{rotoscoping-2006}
have been later introduced by Disney in order to be efficient when working with many single-frame images.
However, these old techniques required artists to illustrate thousands
of pictures to be filmed when close to the end of the production process.

With the appearance and rapid development of computer hardware and software,
traditional animation techniques, such as above, have been slowly replaced
by computer generated graphics and animation. The techniques of key-framing,
morphing, and motion capture~\cite{HB99} have been widely used.
This is the type of animation we are focusing in this work as well as
the use of CG in other areas for documentary production.

Since {\em Walking with Dinosaurs} (1999), a six-part documentary series produced by the BBC
in 1999~\cite{walking-with-dinosaurs-1999}, the style of a traditional documentary
was simulated. More and more documentary films these days include visual
effects and CGI (Computer Generated Images) as a norm.
Comparing to {\em Walking with Dinosaurs}, until the present day
still the most expensive documentary series per minute ever made,
production costs have come down in recent years with the rapid development of
computer hardware and software. Thus, even for those documentary films with low budget,
directors could think about introduce CG animations to their film nowadays.
This new genre of documentary films usually reconstructs the historical
and informational footage, which is not available.

\begin{quote}
{\it ``If you have several visual effects shots that work together to 
tell a story, using them separately earlier on in the documentary 
can increase their impact. ... Earlier in the documentary, 
each shot will work well in sections of the film detailing the different 
aspects of the threat, and then when the entire animation is brought together 
for the climax of the documentary, 
the effect will be even greater.''}~\cite{3d-animation-motion-cgi-documentary}.
\end{quote}

\noindent
Moreover, computer generated graphics and special effects can portray the facts,
and even more advanced than reality, and something called
{\em augmented reality} (AR)~\cite{cost-refs-ar-gi08,arpracticalguide2008,7thingsar}.
It can illustrate very serious and heavy topics with a humorous connotation
in order to attract the audience's attention.
Of course, the visual effects and animation is only one aspect
or role of the vast computer graphics topic.
Obviously, it weights more than traditional animation
for documentary film production.

\subsection{Interaction}

Interactivity is another important element of the emerging documentary film production,
in which other broadcast media cannot not compete with.
The interactivity of the new genre of documentary film enables
the audience to make the decision on what will
be going on in the film. They could participate on what should
be included in the documentary film and in which order.
Today, mass media can be easily used for documentary production
with video clips shot by virtually anyone, uploaded to Youtube
or similar services.
Some documentary-specific online projects allow
user control and interactive documentary making from either
pre-compiled scenes or video clips or user-generated video content
on a particular topic (e.g. weddings, city streets, etc.).

\subsection{Summary}

Thus, how does computer graphics (CG)
contribute to the documentary genre and what is its overall role?
We will summarize the recent research results and examples in the following aspects:

\begin{itemize}

\item Animation
	\begin{itemize}
	\item Computer-animated mock documentaries
	\item Computer-animated documentaries for the Internet
	\end{itemize}

\item Perception
	\begin{itemize}
	\item 2D
	\item 3D, stereoscopic
	\end{itemize}

\item Interaction
	\begin{itemize}
	\item Interactive documentaries and the corresponding research 
	\item Computer game-based documentaries
	\end{itemize}

\item Tools
	\begin{itemize}
	\item Existing software and hardware used to create documentaries
	\end{itemize}
	We mention the tools point, such as Maya, Blender, 3DS Max, XSI~\cite{maya,blender,3dsmax,xsi}
	and others, but we do not go into the details of those tools in this work
	and leaving it to explore in our future work.

\end{itemize}

We will also have a closer look at some of the following documentary films
featuring computer graphics techniques to a various degree that
feature CG-based animation and interaction:
{\em Super Size Me} (2004)~\cite{super-size-me}, a personal experimental
documentary with some computer animation sequence;
{\em Little Voices} (2003)~\cite{little-voices}, a hybrid documentary and computer animation film;
{\em Man With a Movie Camera: The Global Remake} (2007)~\cite{man-with-a-movie-camera-2007}, a very good example to represent the concept of interactivity in documentary;
{\em Escape from Woomera} (2003), a documentary game;
{\em Chicago 10} (2007)~\cite{chicago-10}, a unique and unconventional documentary uses motion-capture animation to portray the Chicago Conspiracy Trial;
{\em Waltz with Bashir} (2008)~\cite{ide-waltz-with-bashir}, an animated documentary film advertised as being the first feature-length animated documentary;
{\em Ryan} (2004)~\cite{ryan-orion-2005,ryan-npar-2004}, a prominent example of an ``all-out'' graphics documentary.

We analyze the impact of the computer graphics on the present-day
documentaries in the two major sub-classes and project the future of the documentary films and TV production,
as well as potential cognitive pattern-recognition-based
interactivity with the documentary scenery might look like in the not-very-distant
future as the professor's responsive hologram in ``I, Robot''~\cite{movie-irobot}
motion picture.

\section{Computer Animated Documentary}

\subsection{CGI and Realism of Documentary}
\label{sect:cgi-and-realism}

\begin{quote}
{\it ``There is no pre-existing reality, no pro-filmic event captured in its occurrence,
an animated film exists only when it is projected.
With no any existence in the world of actuality, the animated film like the partially
dramatized documentary, rely on a kind of artistic re-enactment, depending, in part,
on imaginative rendering as a compensation for the camera's non-presence at the
event.''}~\cite{truth-toons-documentary-1997}
\end{quote}

\noindent
Several types of realism might be identified~\cite{life-reproduced-drawings-2005}
in the following paragraphs extended from \xs{sect:introduction}.
We not only explain their concepts, but also look at computer graphics
role on how enhance theses realism, and which realism types are applicable
for our study.

\paragraph{Visual Realism}~\cite{life-reproduced-drawings-2005} evaluates the extent to which the animated
environment and characters are understood by the audience compared to
the ones from the actual physical world.  Dimensionality and the level of detail (LOD) are
two main aspects of visual realism. Dimensionality refers to the extent that an
illusion of depth is created, whereas LOD describes the extent to which the background
depicts complex particularities of the environment.
3D modeling software, e.g. Maya~\cite{maya}, can not only provide
the advance modeling tools for shaping the objects, but also supply
the advanced rendering techniques for artist to build the very realistic environment
with vivid texture and lighting. Moreover, even an individual without any drawing skills,
can us most of the current 3D modeling software~\cite{maya} to model the objects,
characters, and landscapes~\cite{life-reproduced-drawings-2005}.

The visual realism type is the most applicable from the CG point of view in our study as this
depicting complex particularities of the environment. The visual realism
type is the most applicable from the CG point of view in our study as this
is where it impacts and impresses the audience most.

\paragraph{Aural Realism}~\cite{life-reproduced-drawings-2005} is similar to
the visual realism in the previous paragraph but at the sound level.
It is an important aspect of the realism to enhance the documentary film
perception by the audience (e.g. as in {\em Little Voices} of real
children recorded), and nowadays a lot of it is also computer-generated,
but it is outside of the scope of this work; we mention it for completeness.

\paragraph{Realism of Motion}~\cite{life-reproduced-drawings-2005} contrasts
the extent to the characters moves
and motion in artificial environment and physical world and laws of physics.
Traditional animation relies on persistence of vision and refers to a series
motion illusions resulting from the display of static images in rapid-shown
succession. Artists have to use not only their drawing skills and intuition,
but also posses some knowledge of physics to make the objects behave as if
they are in the real world or close. The motion of the virtual objects
will not convince audiences if no natural laws of physics are applied.
Moreover, drawing the virtual objects move from one frame to another frame
is an inefficient way without functionality provided by software. However,
one of the computer techniques, motion capture is very efficient and accurate
to describe virtual objects motion. It attaches sensors on actors bodies and
records the data for their movements and apply these data to a computer
generated characters. This technique increases the realism of
motion dramatically~\cite{life-reproduced-drawings-2005}.

Additionally, some physics engines combined with computer graphics rendering techniques, e.g. softbody
simulation~\cite{msong-mcthesis-2007,softbody-framework-c3s2e08,adv-rendering-animation-softbody-c3s2e09},
contribute a lot in this realism type by tweaking physical simulation parameters
of the laws, such as gravity, material properties, inertia, one can produce
interesting visual motion outcomes to make a point in a film.

\paragraph{Narrative and Character Realism}~\cite{life-reproduced-drawings-2005} attempts
to make audience believe the fictitious events and characters of the animated
film actually exist. For example, in order to portray the animated character vividly,
artists use the squash-and-stretch method exaggerated for soft parts
of the character in traditional animation techniques~\cite{life-reproduced-drawings-2005}.
However, it is a very time consuming procedure.
Today's computer graphics software often provide a group of functionality and library,
such as hair, skin, clothes animation, skeleton animation in order to simplify
artists' work and achieve more realistic results~\cite{life-reproduced-drawings-2005}.

\paragraph{Social Realism}~\cite{life-reproduced-drawings-2005} makes audience believe
that the event take place in the fictitious animated world is as complex and diverse
as the real world. This concept applies both on traditional animation and
computer animation. In order to achieve social realism, artists not only rely on other
visual, character, motion realism, but also count on the writing
of the documentary production~\cite{life-reproduced-drawings-2005}.

\paragraph{Psychological Realism} was first time brought up by Chris Landreth,
the director of animated short documentary. It does not consider the physical
based motion as the priority, nor use some techniques, such as rotoscoping
and motion capture. Instead, it uses an original, personal, hand animated
three-dimensional world which Landreth calls ``psychological realism''.
Using a technique called psychological realism, the movie shows
the emotions of the characters in a way never seen before.

\subsection{Types of Visual Effects and Animation in Documentaries}
\label{sect:visual-effects-types}

Christian Darkin~\cite{3d-animation-motion-cgi-documentary} categorized
the types of computer graphics and the corresponding techniques
used in documentary film as following in several categories that
we recite below to complement our study.

\paragraph{Explanation Graphics} according to Darkin~\cite{3d-animation-motion-cgi-documentary},
is a quite acceptable means to provide explanatory notes, ideas,
as well as information when the available footage
cannot portray it sufficiently well. The CG-animated explanatory
supplements can be very exciting and creative; 2D or 3D; text, cartoon,
moving characters -- it is
up to the director, animator, artists, and their creativity,
resources available, and the corresponding needs of what to
portray. Such as CG type can arguably be fitting with any
documentary style~\cite{3d-animation-motion-cgi-documentary}.

\paragraph{Animation for Color Shots} -- is a general CG mechanism
that is applicable an relevant to many types of documentary
film production, especially if at some point there is more
narration material than footage~\cite{3d-animation-motion-cgi-documentary}.
Darkin further gives good examples of such color shots, such as
a 20-second 3D animation where the virtual camera rushes
through a bloodstream of a patient and blood cells and others
``fly'' past can easily be used in many medical-related documentaries
and be ``on topic'' and not boring. Similarly, in the crime-related
documentary, one can animate a ``fly-through'' a CG-generated
building where the crime happened while narration is running and
prior to when the real footage begins~\cite{3d-animation-motion-cgi-documentary}.

\paragraph{Visual Effects Reconstructions} as opposed to the explanation graphics,
are required in the absence of footage for the most prominent and necessary
events to portray in the documentary film that could not have been possibly
shot, physically inaccessible (the scale of cellular biology or the Universe),
or too far in the past, e.g. the assault on Baghdad,
dinosaurs, an assassination, the Big Bang, or if needed to show the
fat molecules getting from a burger to one's thighs, the 3D animated
CG reconstruction can greatly help to portray such
events in a documentary film~\cite{3d-animation-motion-cgi-documentary}.

\paragraph{Text and Title Animation}
is commonly employed to bring up documentary titles,
scene announcements, some short on-screen paragraphs
or questions raised by the maker, and subtitles.
An introduction to the film, its topics, and ideas are
good candidates for this type of animation in order to
set the tone the documentary film is to proceed with. Some
of the Motion Graphics techniques mentioned earlier can
be very well applied to text as well here as the CG
techniques are typically common across the board, except
the text animation is optional and can be just
a still~\cite{3d-animation-motion-cgi-documentary}.

\subsection{Examples of Computer Animated Mock Documentaries}
\label{sect:example-mock-documentary}

\paragraph{{\em Super Size Me} (2004)} by \cite{super-size-me} is a small budget documentary film,
produced by an independent filmmaker Morgan Spurlock. In this film, he did a test on himself that
he had to only eat McDonalds three meals a day for 30 days~\cite{super-size-me}.
The film is very fast pace and
full of sense of humor. It is ``in-between'' animated documentary, which contains partially
animated graphics and scenes according to our classification on animated documentary film.
It combines computer generated graphs, charts, and animation, with the tune in humor,
which is the best way to get an audience interested in topics.

The simple graphics captured from the film show in \xf{fig:super-size-me-1},
is 100 times more visually vivid for audience to understand the information about
how fast food would cause a 10 year old and a 14 year old girl to gain so much weight.
Another example of animation used in this film shown
in \xf{fig:super-size-me-2} exaggerates how McDonalds make the chicken nuggets from
``unusually large breast chicken''.

\begin{figure*}[ht]
\hrule\vskip4pt
\begin{center}
	\subfigure[Example 1]
	{\label{fig:super-size-me-1}
	\includegraphics[height=2.0in]{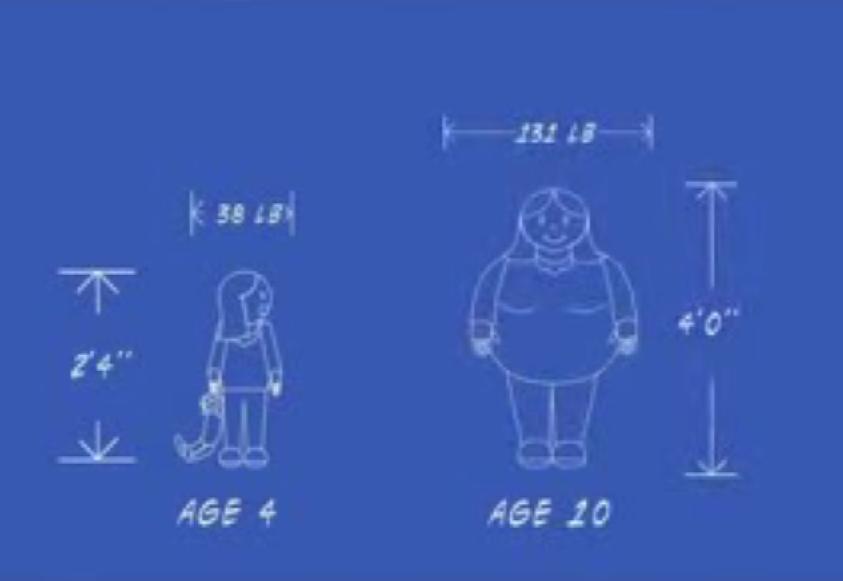}}
	\hspace{.3in}
	\subfigure[Example 2]
	{\label{fig:super-size-me-2}
	\includegraphics[height=2.0in]{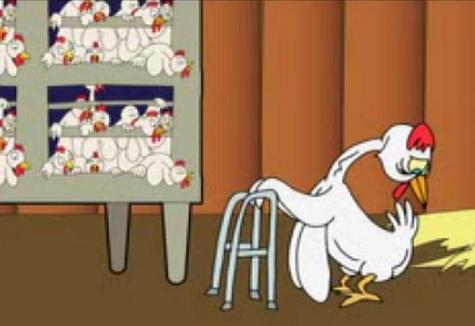}}
\caption{{\em Super Size Me} (2004) Example Screenshots}
\label{fig:super-size-me-screenshots}
\end{center}
\hrule\vskip4pt
\end{figure*}

When Morgan Spurlock was asked of he used ``a lot of video games, a lot of computer graphics
and cell animation'' and if it is the way to make his points
``more accessible to the mainstream'', he replied,

\begin{quote}{\it ``Definitely. Absolutely. One of my beliefs as a filmmaker is that if you can make somebody
laugh, you can make them listen. With laughter, you can get somebody's guard down,
you can open them up to listening to you.
They don't feel like they're being preached to or talked down to.
I think it helps, it makes really hard to understand information
a little more accessible and palatable. And at the end of the day,
it makes a movie a little more fun. It doesn't feel so heavy handed.''}
\end{quote}

\paragraph{{\em Little Voices} (2003)} by~\cite{little-voices} is a uniquely done
mix of of techniques making the documentary what the scholars call ``hybrid'' that
included computer animation into the film. The film's director, Jairo Eduardo Carrillo
made a number of interviews of displaced children in Colombia's capital, Bogota during the
Colombian Civil War. The core theme of the film is runs through the real stories
told by the real children in their own voices, but the stories themselves were
illustrated initially by the children's drawings and paintings of the scenes
they were describing. Then Carrillo took those 2D drawings of characters, scenery,
etc. made by the children and turned them into the animated 3D CG models.
A combination of the children's art, computer animation, virtual and augmented reality
techniques together with the children's voices create an impressive environment
and an art piece for the audience that not only engaging, but also preserving
the charm, integrity, and energy of the original art work of the children in their
3D animated counterparts~\cite{little-voices}. Following the creation of this
hybrid documentary film, Carrillo further created an educational game for displaced
children, with the same title {\em Little Voices}
while he was staying during his residency at the Banff New Media Institute (BNMI).
In \xf{fig:little-voices-screenshots} are examples of a drawing,
then a 3D-redone scene, and the photograph of two children participants.

\begin{figure*}[ht]
\hrule\vskip4pt
\begin{center}
	\subfigure[Example 1]
	{\label{fig:littlevoices-1}
	\includegraphics[height=1.1in]{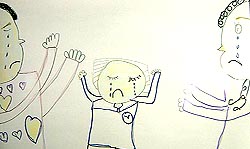}}
	\hspace{.3in}
	\subfigure[Example 2]
	{\label{fig:littlevoices-2}
	\includegraphics[width=1.5in]{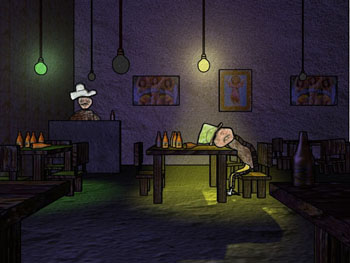}}
	\hspace{.3in}
	\subfigure[Example 3]
	{\label{fig:littlevoices-3}
	\includegraphics[width=1.5in]{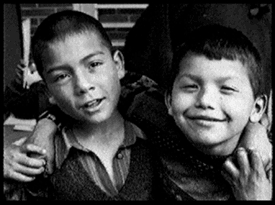}}
\caption{{\em Little Voices} (2003) Screenshot Examples}
\label{fig:little-voices-screenshots}
\end{center}
\hrule\vskip4pt
\end{figure*}

\begin{quote}
{\it ``There is a certain poignancy in the Little Voices project, that can be found in the
intimate self-portraits drawn by the young people of Bogota recontextualized
into a large-scale animated world that is both compelling and deeply disturbing,''}
says BNMI director Susan Kennard. {\it ``Carrillo draws our attention to this
juxtaposition through a representation of place that is both real and unreal.''}~\cite{speaking-through-little-voices}
\end{quote}

\noindent
{\em Born Under Fire} (2008) is another computer animated CGI
documentary by Carrillo that follows the similar style as the
{\em Little Voices}, but this time around, it is based on the
interviews with and drawings by the new generation of children who
were 8 to 13 years old at the time of the interview, and have grown up
in midst of violence and chaos in Colombia~\cite{born-under-fire}.

\paragraph{{\em Chicago 10} (2007)} is the documentary film~\cite{chicago-10} about
the Chicago Seven incident. This is known as a very good and commended example where
the visual effects reconstruction type of graphics used for animating
the missing footage of the court room scenes of the Chicago Conspiracy Trial
all the way back in 1968 where the animation is nicely blended with some available
footage archives back from 1968 in order to accent the development
of the story and its emotion more sharply~\cite{chicago-10}.
Some example screehshots from the animation are in \xf{fig:chicago-10-screenshots}.

\begin{figure*}[ht]
\hrule\vskip4pt
\begin{center}
	\subfigure[Example 1]
	{\label{fig:chicago10-1}
	\includegraphics[height=2.0in]{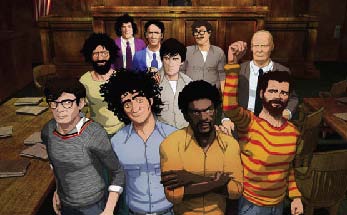}}
	\hspace{.3in}
	\subfigure[Example 2]
	{\label{fig:chicago10-2}
	\includegraphics[height=2.0in]{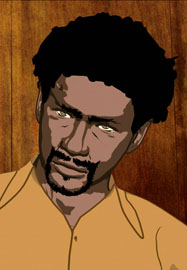}}
\caption{{\em Chicago 10} Example Screenshots}
\label{fig:chicago-10-screenshots}
\end{center}
\hrule\vskip4pt
\end{figure*}

\paragraph{{\em Ryan} (2005)}~\cite{ryan-interview-2005,ryan-orion-2005} won Oscar 77th
annual Academy Awards for Best Animated Short Film.
In this documentary, based on a period of life of Canadian animator Ryan Larkin,
the audience perceives the voice of Ryan and the surrounding people, but
3D CGI characters visualizing Ryan and the others appear a bit strange,
twisted, see-through, sometimes broken and disembodied, which are humorous
or disturbing at times~\cite{ryan-interview-2005}.
While the rendering of the CGI scenes in {\em Ryan} is non-photorealistic~\cite{ryan-npar-2004},
the 3D characters and the virtual environment are very detailed and make an
impression of being very realistic despite the fact that this documentary was created
not with the use of rotoscoping or motion capture techniques presented earlier,
but rather by using a 3D modeling software~\cite{maya} with some extra modifications
and plug-ins to enable the non-photorealistic rendering and mixed perspective and
non-linear projection~\cite{ryan-npar-2004}. Scholars classify this as an
autobiographical documentary, but non-traditional, because the whole film,
even the interviews were turned 100\% into 3D computer graphics scenes
instead of being filmed by a live-action camera~\cite{ryan-interview-2005}.
Director of this film, Chris Landreth,
says that after he learned Ryan Larkin's story:

\begin{quote}
{\it ``There's a lot wrapped up in that, as far as a great story to tell, and I wanted to tell that in a
way that was as powerful as possible. Making an animation out of
a documentary was the best way, in my opinion, to do that.''}~\cite{ryan-interview-2005}
\end{quote}

\noindent
Examples of the the {\em Ryan}'s CG screenshots are in \xf{fig:ryan-screenshots}.

\begin{figure*}[ht]
\hrule\vskip4pt
\begin{center}
	\subfigure[Example 1]
	{\label{fig:ryan}
	\includegraphics[height=2.0in]{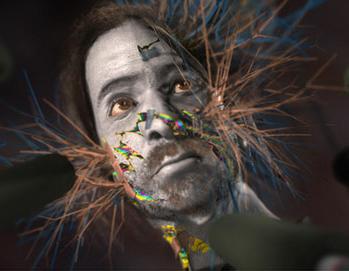}}
	\hspace{.3in}
	\subfigure[Example 2]
	{\label{fig:ryan2}
	\includegraphics[height=2.0in]{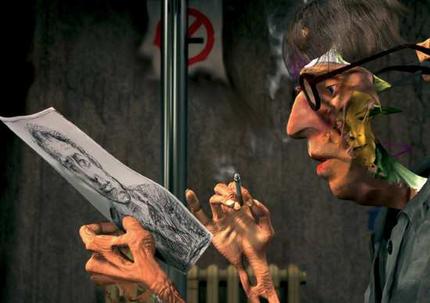}}
\caption{{\em Ryan} Example Schreenshots}
\label{fig:ryan-screenshots}
\end{center}
\hrule\vskip4pt
\end{figure*}

\paragraph{{\em Waltz with Bashir} (2008)}~\cite{ide-waltz-with-bashir} is a CGI-animated
documentary film, which was advertised as being the first
feature-length animated documentary except the short part in the ending
that was featuring the real documented results of the Sabra and Shatila
found in an archived footage from the news at the time.
Overall, it took four years for the director to complete the film,
which is ``a combination of Flash animation, classic animation, and 3D''~\cite{sony-waltz-with-bashir}.
Two example screenshots from the documentary are in \xf{fig:waltz-bashir-screenshots}.

\begin{figure*}[ht]
\hrule\vskip4pt
\begin{center}
	\subfigure[Example 1]
	{\label{fig:waltz}
	\includegraphics[height=2.0in]{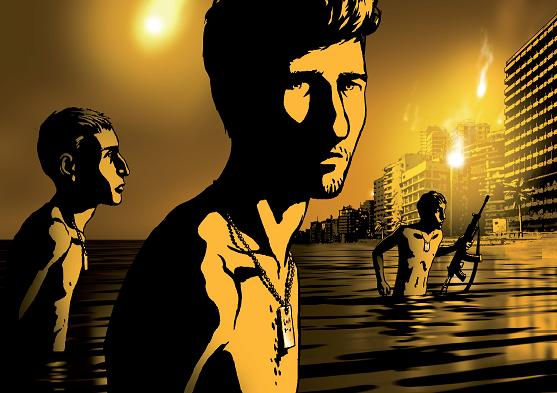}}
	\hspace{.3in}
	\subfigure[Example 2]
	{\label{fig:waltz2}
	\includegraphics[height=2.0in]{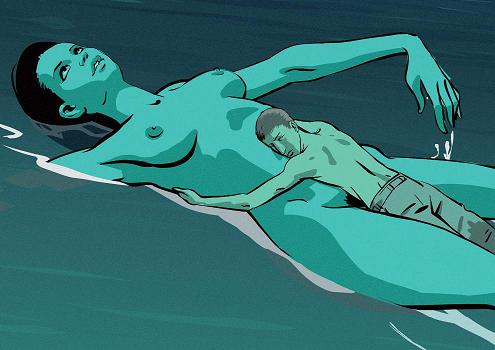}}
\caption{Waltz with Bashir Screenshot Examples}
\label{fig:waltz-bashir-screenshots}
\end{center}
\hrule\vskip4pt
\end{figure*}

In this documentary Israeli director Ari Folman attempts to reconstruct
his missing memories from his time as a soldier in the 1982 Lebanon War by using
the animation.
The techniques used in the film can be confused with the {\em rotoscoping},
mentioned earlier -- an animation style that uses drawings over live
footage -- but this is in reality a combination of Adobe Flash scenes and
classic animation techniques~\cite{ide-waltz-with-bashir,sony-waltz-with-bashir}.
Folman by using the freedoms that animation provides to take the
file into scenes that could not have been possible to shoot in the
traditional way. It also impacts the audience's preconceptions
of cartoons always belonging to the realm of narrative filmmaking
and attempting to emphasize where the line between the
fiction and reality lies~\cite{ide-waltz-with-bashir,sony-waltz-with-bashir}.

\begin{quote}{\it ``For a few years, I had
the basic idea for the film in my mind but I was not happy at all to do it in real life video. How would that
have looked like? A middleaged man being interviewed against a black background, telling stories that
happened 25 years ago, without any archival footage to support them. That would have been SO BORING!
Then I figured out it could be done only in animation with fantastic drawings. War is so surreal, and memory
is so tricky that I thought I'd better go all along the memory journey with the help of very fine illustrators.
The animation, unique with its dark hues representing the overall feel of the film, uses a unique style
invented by Yoni Goodman at the Bridgit Folman Film Gang studio in Israel.''}~\cite{sony-waltz-with-bashir}
\end{quote}

\noindent
This opens up the tools and techniques used in the production process to complement
the missing footage. Despite the documentary being mostly animated, the Folman's
story was told undistracted and still impacting the audience. On the team the
animator was Yoni Goodman, who produced the majority of the types of styles of
vivid and stunning animation, sometimes hand-drawn, he did not obscure the main
point of the story~\cite{drawing-from-memory-2009}.
First, each his drawing was sliced into hundreds of pieces that were moved
in relation to one another in order to create the movement. Then, the film was
preliminary shot in a sound studio as a 90-minute video, which was then transferred
to a storyboard. From there about 2300 original illustrations were drawn
with respect to the the storyboard. All those illustrations, which together
eventually formed the actual film scenes, were composed using the aforementioned Flash animation,
classic animation, and other 3D technologies~\cite{israeli-filmmakers-2008}.

\section{Interactive Documentary}

Accompanied with the development of digital media and computer-based technologies,
new forms of documentary are challenging the traditional ones everyday.
Except the CGI documentary we discussed in previous section, interactive documentary
is another new type directly related to interactive computer graphics~\cite{ea03},
which had been widely used in video game programming. Only since the 21st century,
interactive computer graphics has been slowly introduced to documentary production.
The related projects are mostly web based to enable user to not only view but also
participate the making of documentary film. Some pioneers of
``Interactive Online Documentary''~\cite{interactive-online-documentary} such as
Australian Film Commission (AFC)~\cite{afc} and Australian Broadcast Corporation (ABC)~\cite{abc}
collaborated with local Australian filmmakers and digital media artists to work on
these new documentary projects.

\subsection{Participation}

{\em Man With a Movie Camera: The Global Remake} (2007) is the best example to represent
the concept of interactivity of documentary~\cite{man-with-a-movie-camera-2007}.
It is inspired by Vertov's {\em Man With A Movie Camera} made in 1929.
The original documentary film records the progression of one full day synthesizing
footage shot in Moscow, Riga, and Kiev. It begins with titles that declare it
``an experiment in the cinematic communication of visible events without
he aid of intertitles, without the aid of a scenario,
without the aid of theater.''~\cite{perry-bard-interview}
Vertov's footage was shot in the industrial landscape of the 20's.
What images translate the world today? The current project is a participatory
video shot by people around the world who are invited to record images
interpreting the original script of Vertov's film and upload them to the website.
Anyone can upload footage and contribute as part of a worldwide montage.
The artist Perry Bard says,

\begin{quote}
{\it ``Vertov's 1929 film is a great point of departure
for the Internet because it has so many dimensions from the documentary to the
performative to the effects along with its use of an archive which translates
to a database and it's a film within a film. It was shot in three different cities,
going global was obvious and the rhythm is very contemporary, there's no shot in
the film longer than twenty seconds. It seemed like a perfect vehicle for
global input and in keeping with Vertov's intentions as a filmmaker.''}~\cite{perry-bard-interview}
\end{quote}

\subsection{Perception, Dimensionality, Interaction}

\subsubsection{2D}

{\em Jerome B. Wiesner (JBW) (1915-1994): A Random Walk through the 20th Century}, which was developed in 1996 by
Golrianna Davenport's MIT Media Lab on interactive cinema, is part of the
``evolving documentary'' genre~\cite{mit-2004}.
This ``hyper-portrait'' introduces the audience to a remarkable man whose life centered on science,
government, education and issues of cultural humanism~\cite{mit-2004}. In this hyper portrait (which runs on
the World Wide Web), audience is invited to explore the 20th century through
an extensible collection of stories about and recollections by the central figure.
Audience who knew JBW are also invited to share a memorable story
with the growing society of audience~\cite{mit-2004}.

Another similar project developed by the MIT Media Lab is {\em Traveling in Zagori}.
Through visual and textual snapshots of the landscape, architecture and people,
the audience is invited to construct a story about a local legend while
discovering aspects of Zagori's extraordinary history and legacy~\cite{mit-2004}.

We would like to group these type of project developed by MIT research
group~\cite{mit-2004} as 2D interactive documentary. Usually, there are some
pre-filmed video footage and audience could decide which one to watch,
and regroup them into different sequence in order to make their own
stories according to their interests and point of view.
This new ``Evolving Documentary'' will increase audience's
interests to participate and interact with the film rather
than being a viewer with no input.

\subsubsection{3D}

``Interactive Cinema'' or ``Interactive Fiction'' are not very new terms,
which are normally taken to be 3D computer games, typically of the adventure
or quest type. The first interactive documentary, {\em The Unexplained},
says FlagTower, has named this concept the Interactive Documentary,
a title which reflects the televisual appeal of the style of production.

3D Interactive Documentary projects the interrelated sequence of moving
images on a screen and give user a view of a word including agents,
``props'' and scenery, organized in one or more locations~\cite{dankert-erik-2000}.
In Dankert and Wille's project, they used a single computer screen as projection,
others could apply `Virtual Reality' to the production.

Another 3D aspect -- stereoscopy -- that comes with the new
stereoscopic movies documenting potential life of animals
millenia ago, or the animals now in the see or simply
documenting a concert in 3D, such as done by IMAX for
their animations and stereo-enhanced motion pictures
such as {\em Sea Monsters} (2008) and {\em U2} (2008) concert. The audience
in this case, while swimmingly passive and watching,
interacts with the show through the stereoscopic illusion
their eyes receive as a part of the show enhancing the
perception and impression of the show. One can argue
it is one-way interactive, but nonetheless should not
be neglected. Needless to say, the stereoscopic shows
can also be made interactive, and even more so in the
future.

\section{Future Work and Conclusion}

According to the recent work by Landesman~\cite{ohad-landesman-2008},
the documentary film overall in the past years has been experiencing
a notion of formal change from traditional ``observation and omniscient narration''
in terms of being less strict in the need to objectively portray
the material. More and more the documentary film has been embracing
the paradigm switch to performance rather than recording of an observation,
some subjective rather than objective aspects and even fiction; equally
no longer requiring to be as certain and complete as possible
in the argumentation in the film instead of just showing dry
factual knowledge~\cite{ohad-landesman-2008}. CGI techniques
are here to help with the emerging trends of documentary film concepts
and production Landesman described.

Today, CG and animation are more advanced than traditional animation
according to the interactive, physical based (mimicking live-action camera),
more effective visual result, and less burden on the financial/budget concerns.

Additionally, in the past, the autobiographic documentary film type
was often constrained to for film makers themselves only as the
ones possessing the knowledge, resources, and equipment to do so.
With the wider accessibility of
digital content and high computing power these days, autobiographic
documentaries are now accessible to virtually any household.

Furthermore, the notion of script writing in terms of screen writing -- i.e.
the writing for the film screen or television screen for the new genre of documentary films
is an emerging, new research area requiring more work.

Moreover, teaching in the research and application
of ``CGI for documentary films'' is also an important
topic that is going to be inevitably introduced into
the core of Film Studies for scholarly research
and film production industry.

\paragraph{Position.}

I can not agree with those who deny the value and important
role of computer graphics in documentary production.
An extreme example of a such a point of view is William Moritz's:

\begin{quote}
{\it ``No animation film that is not non-objective and/or non-linear can
really qualify as true animation, since the conventional linear representational
story film has long since been far better done in live-action.}~\cite{william-moritz-1988}
\end{quote}

In my opinion, CGI documentary is a unique form of documentary which not only gives
the filmmakers and media artists enough space for the their creativity,
but also bring audience a wonderful experience. If ``truth-value'' and ``realism''
are two opposing terms to traditional reconstructed documentary film,
they do not conflict in animated documentary.
It makes audience aware that what they are watching are not real
live-action footage, but reconstructed based on artists creation.
Some traditional documentary film, some of the reconstructed scenes
could really fool the audience by the perfect ``realism''. Audience
could be moved by the reconstructed scenes simply because they believe
what's being shown is a real event. Imagine their disappointment, when
they learn it was faked to impress them. However, if they aware that
what they see is not real-footage people, real-footage event, how could
they appreciate the truth-value of the film? In animated documentary,
we do not have such worry because it is very honest to the audience
that what they see are not a ``reality show'' but a creation and art of reconstruction.
I believe this aspect is most precious ``moral character'' of animated
documentary film, which is best element in evaluation of truth-value.
Compared to traditional documentary film, the more realism of the
reconstructed scenes to make the audience believe in it, the more misleading
to the audience; the more increase of the realism in animated documentary,
the more audience could appreciate the art value of the film,
and review some of the historical events as if they were there.

It is not a coincidence that some hold similar position as I do.

\begin{quote}
{\it ``The relationship between reality, documentary and animation could
therefore be described as `creating the real'... if we can argue that live
action documentary attempts to `claim' the real by virtue of its immense mimetic
potential, and the conventions of correspondence that have accrued to this mode
of representation, then it is similarly persuasive to suggest that animated
documentaries `create'.''}~\cite{paul-ward-2008}
\end{quote}

\begin{quote}
{\it ``Animation can be regarded as a genuine source of documentation, where new technologies, political stands, and economical status contribute greatly to its formation. Even if certain works are subjectively made to reinforce certain point of views they are still a product of their time where subconscious and conscious choices reveal a truth about its society.''}~\cite{alan-kininsberg-2007}
\end{quote}

\indent
A Chinese proverb says,
``A picture is worth thousands words.'' I think that is why some of the
old silent film are still so powerful today. At some point the language
becomes so plain and useless compared to the images.
Some feel there is the need to use animation in their documentary film production,
such as Morgan Spurlock (the director of {\em Super Size Me}), Chris Landreth (the director of {\em Ryan}),
and Ari Folman (the director of {\em Waltz with Bashir}) who explain the reasons
why their documentary films fully or partially contain computer
generated graphics. Not only because they have to use this technique
to fill in the gaps of missing footage,
but also because they fell that is a freer and more proper way to meet
their creativity. From the truth-value aspect,
the CGI images are closer to the truth than ``talking head'' interview,
especially reconstructed scene and not existing scene.

I believe CGI documentary is the future direction of documentary production.
In the ancient times, human beings did not invent a camera yet, so
the only way to capture the images was drawing. As many years ago when
we did not have the Internet and telephone, our only communication
was by a regular mail delivered by a mailman. When we had no sources
to do more, we had to stay with whatever techniques available. Today,
now, the rapid development of technology, one can use all the available
means to serve the creativity. Documentary production is not an exception.
It has to adapt to the new challenges.

For the younger generation audience, who grow up with the Internet,
mobile phones and different sorts of new and portable media,
a slow pace with ``talking head'' type of documentary
will simply be too boring for them.

For the older audience, if they accept the computer technology,
they will adapt to the same way to new genre of documentary film.
They will be aware that is the development of not only technology,
but also of the whole society and mentality. They will understand
that animated documentary is just another, perhaps more open, form of documentary
film and animation is just another approach for the artists and filmmakers
to create their projects.

It is my position~\cite{haptics-cinema-future-grapp09} that
there inevitably will be a new ``language'' for documentary,
tools, and techniques, accessible to mass media.
The process of script writing for formal documentaries is
bound to change to account for the technological advances
or entirely disappear for the mass media volume productions
and autobiographical, ad-hoc, documentaries.
The use of webcasts, podcasts, interactive and Internet-based
television, Youtube-worth of archival data
combined with the use of a virtual camera, clipping, super-resolution
techniques are already impacting the documentary production
redefining the documentary landscape, what I'd call the {\em docu-scape}.

The human-computer interface (HCI) design, along with
haptic~\cite{wiki:haptic-techonology,haptics-large-workspaces},
devices allowing back propagation and physical feedback
from the virtual to the real will enhance the interactivity
of the future documentary adding a notch of physical touch
to the interaction of the audience with the film's virtual
fabric. Further introducing stereoscopic effects,
and virtual and augmented realities might even make a feeling
the the documented events, educational programs, and what
not is happening like in the Matrix~\cite{matrix-movie}
or, at the very least, like 3D IMAX movies.

Interactivity, plus virtual and augmented realities, is
definitively the future of documentaries. Dynamic,
physical-based real-time documentation of events,
reports, logs, anything, as well as integrated
simulations, can be put into a small device such
as a holocron used by the Jedi or the Sith
from the {\em Star Wars}, or the killed professor
from {\em I, Robot}~\cite{movie-irobot} to store the interactive knowledge
and an image of themselves (they may not even open
to everyone or wait for the right questions to be asked,
like the police officer was trying to determine what
happened to the dead professor in {\em I, Robot}, who pre-recorded
his interactive hologram that was designed to answer
the right questions). Such holocrons could be the
future media in the not so distant future to record
and document the events and the knowledge.

\clearpage
\bibliography{cg-documentary}
\bibliographystyle{apalike}

\end{document}